\documentclass[11pt]{article}
\usepackage{times}
\usepackage{geometry}
\usepackage[utf8]{inputenc}
\usepackage{enumitem,amssymb}
\usepackage{ragged2e}
\usepackage{graphicx}
\usepackage{comment}
\usepackage{multicol}
\usepackage[usenames]{xcolor} 

\usepackage[left]{lineno}

\definecolor{xlinkcolor}{cmyk}{1,1,0,0}
\usepackage{url}
\usepackage[
 colorlinks=true,    
 linkcolor=xlinkcolor,     
 citecolor=xlinkcolor,     
 filecolor=xlinkcolor,  
 urlcolor=xlinkcolor,      
 final=true
]{hyperref}
\usepackage[super,sort&compress]{natbib}
\usepackage{enumitem}
\setenumerate{itemsep=0mm}

\begin{document}
\begin{raggedright} 
\huge
Snowmass2021 - Letter of Interest \hfill \\[+1em]
\textit{New opportunities at the photon energy frontier} \hfill \\[+1em]
\end{raggedright}

\normalsize



\noindent {\large \bf Coordinators of this LoI:}
Spencer Klein (LBNL)\footnote{srklein@lbl.gov} and  
Daniel Tapia Takaki (U. Kansas) \footnote{Daniel.Tapia.Takaki@cern.ch}

\noindent {\large \bf Authors:}
Jaroslav Adam$^{9}$,
Christine Aidala$^{40}$,
Aaron Angerami$^{3}$,
Benjamin Audurier$^{47}$,
Carlos Bertulani$^{17}$,
Christian Bierlich$^{24}$,
Boris Blok$^{35}$,
James Daniel Brandenburg$^{9}$,
Stanley Brodsky$^{34}$,
Aleksandr Bylinkin$^{2}$,
Veronica Canoa Roman$^{42}$,
Francesco Giovanni Celiberto$^{52}$,
Jan Cepila$^{0}$,
Grigorios Chachamis$^{46}$,
Brian Cole$^{22}$,
Guillermo Contreras$^{0}$,
David d'Enterria$^{14}$,
Adrian Dumitru$^{28}$,
Arturo Fernández Téllez$^{20}$,
Leonid Frankfurt$^{10,50}$,
Maria Beatriz Gay Ducati$^{19}$,
Frank Geurts$^{23}$,
Gustavo Gil da Silveira$^{11}$,
Francesco Giuli$^{26}$,
Victor P. Goncalves$^{16}$
Iwona Grabowska-Bold$^{5}$,
Vadim Guzey$^{12}$,
Lucian Harland-Lang$^{32}$
Martin Hentschinski$^{29}$,
T. J. Hobbs$^{25}$,
Jamal Jalilian-Marian$^{28}$
Valery A.  Khoze$^{15}$,
Yongsun Kim$^{36}$,
Spencer R. Klein$^{1}$,
Simon Knapen$^{21}$,
Mariola K{\l}usek-Gawenda$^{48}$,
Michal Krelina$^{0}$,
Evgeny Kryshen$^{12}$,
Tuomas Lappi$^{38}$,
Constantin Loizides$^{7}$,
Agnieszka Luszczak$^{44}$,
Magno Machado$^{39}$,
Heikki M\"{a}ntysaari$^{38}$,
Daniel Martins$^{7}$,
Ronan McNulty$^{45}$,
Michael Murray$^{2}$,
Jan Nemchik$^{0}$,
Jacquelyn Noronha-Hostler$^{33}$,
Joakim Nystrand$^{6}$,
Alessandro Papa$^{51}$,
Bernard Pire$^{37}$,
Mateusz Ploskon$^{1}$
Marius Przybycien$^{5}$,
John P. Ralston$^{2}$,
Patricia Rebello Teles$^{18}$ 
Christophe Royon$^{2}$,
Björn Schenke$^{9}$,
William Schmidke$^{9}$,
Janet Seger$^{8}$,
Anna Stasto$^{10}$,
Peter Steinberg$^{9}$,
Mark Strikman$^{10}$,
Antoni Szczurek$^{48}$,
Lech Szymanowski$^{31}$,
Daniel Tapia Takaki$^{2}$,
Ralf Ulrich$^{49}$,
Orlando Villalobos Baillie$^{41}$, 
Ramona Vogt$^{3,4}$,
Samuel Wallon$^{30}$,
Michael Winn$^{43}$,
Keping Xie$^{27}$,
Zhangbu Xu$^{9}$,
Shuai Yang$^{23}$,
Mikhail Zhalov$^{12}$, and
Jian Zhou$^{13}$\\
(0)~Faculty of Nuclear Sciences and Physical Engineering, Czech Technical University in Prague, Prague, Czech Republic; \\
(1)~Nuclear Science Division, Lawrence Berkeley National Laboratory, Berkeley CA; \\
(2)~Department of Physics and Astronomy, The University of Kansas, Lawrence KS; \\
(3)~Lawrence Livermore National Laboratory, Livermore CA;\\
(4)~Department of Physics, University of California, Davis CA;\\
(5)~AGH University of Science and Technology, Cracow, Poland;\\
(6)~Dept. of Physics and Technology, University of Bergen, Bergen, Norway;\\
(7)~Oak Ridge National Laboratory, Oak Ridge, TN;\\
(8)~Physics Department, Creighton University, Omaha, NE;\\
(9~Brookhaven National Laboratory, Upton, NY;\\
(10)~Dept. of Physics, Pennsylvania State\\ University, University Park, PA;\\
(11) UFRGS and UERJ, Brazil;\\
(12)~Petersburg Nuclear Physics Institute of National Research Centre "Kurchatov Institute;\\
(13)~Shandong University(Qingdao), China;\\
(14)~CERN, Geneva, Switzerland;\\
(15)~IPPP, Durham University, Durham, UK;\\
(16)~Universidade Federal de Pelotas, Pelotas, Brazil; \\
(17)~Department of Physics and Astronomy, Texas A\&M University-Commerce, Commerce, TX;\\
(18)~Universidade Federal do Rio de Janeiro, UFRJ, Rio de Janeiro, Brazil;\\
(19)~UFRGS (Universidade Federal do Rio Grande do Sul), Brazil\\
(20)~Benemérita Universidad Autónoma de Puebla, Puebla, México;\\
(21)~Physics Division, Lawrence Berkeley National Laboratory, Berkeley CA;\\
(22)~Dept. of Physics, Columbia University, New York, NY, 10027;\\
(23)~Department of Physics and Astronomy, Rice University, Houston, TX, 77005;\\
(24)~Department of Astronomy and Theoretical Physics, Lund University, Lund, Sweden;\\ 
(25)~Department of Physics, Southern Methodist University and JLab EIC Center;\\
(26)~Unversity and INFN Rome 2, Via della Ricerca Scientifica, 1 – 00133 Roma;\\
(27)~PITT PACC, University of Pittsburgh, Pittsburgh, PA 15260;\\
(28)~Baruch College, CUNY, New York, NY 10010;\\
(29)~Departamento de Actuaria, F\'isica y Matem\'aticas, Universidad de las Am\'ericas Puebla,  Ex-Hacienda Santa Catarina Martir S/N,  San Andr\'es Cholula 72820 Puebla, Mexico;\\
(30)~Université Paris-Saclay, CNRS/IN2P3, IJCLab, 91405 Orsay, France;\\
(31)~National Centre for Nuclear Research, Warsaw, Poland;\\
(32)~Department of Physics, University of Oxford, Oxford, UK;\\
(33)~Illinois Center for Advanced Studies of the Universe, Department of Physics, University of Illinois at Urbana-Champaign, Urbana, IL 61801, USA;\\
(34)~SLAC National Accelerator Laboratory, Stanford University, Stanford, CA;\\
(35)~Physics Department, Technion, Haifa, Israel;\\
(36)~Sejong University, Seoul, South Korea;\\
(37)~CPHT, CNRS, Ecole polytechnique, I.P. Paris, 91128 Palaiseau, France;\\
(38)~Department of Physics, University of Jyväskylä, P.O. Box 35, 40014 University of Jyväskylä, Finland    and   Helsinki Institute of Physics, P.O. Box 64, 00014 University of Helsinki, Finland;\\
(39)~Institute of Physics - UFRGS, Brazil;\\
(40)~Physics Department, University of Michigan, Ann Arbor, MI 48109;\\
(41)~School of Physics and Astronomy, The University of Birmingham, Edgbaston, Birmingham, UK. B15 2TT;\\
(42)~Department of Physics and Astronomy, Stony Brook University, Stony Brook, NY, 11794;\\
(43)~Université Paris-Saclay Centre d’Etudes de Saclay (CEA), IRFU, Départment de Physique Nucléaire(DPhN), Saclay, France;\\
(44)~Cracow University of Technology, Cracow, Poland;\\
(45)~University College Dublin, Dublin, Ireland;\\
(46)~LIP,  Av. Prof. Gama Pinto  2,  P-1649-003 Lisboa,  Portugal;\\
(47)~Laboratoire Leprince-Ringuet, Palaiseau, France;\\
(48)~Institute of Nuclear Physics Polish Academy of Sciences, PL-31342 Krakow, Poland;\\
(49)~KIT, Karlsruhe Institute of Technology, 76131 Karlsruhe, Germany;\\
(50)~Sackler School of Exact Sciences, Tel Aviv University, Tel Aviv, 69978, Israel\\
(51)~Universit\`{a} della Calabria and INFN-Cosenza, Italy\\
(52)~Universit\`{a} degli Studi di Pavia and INFN, Sezione di Pavia, Italy\\

\noindent {\large \bf Abstract:}
\noindent Ultra-peripheral collisions (UPCs) involving heavy ions and protons are the energy frontier for photon-mediated interactions.  UPC photons can be used for many purposes, including probing low-$x$ gluons via photoproduction of dijets and vector mesons, probes of beyond-standard-model processes, such as those enabled by light-by-light scattering, and studies of two-photon production of the Higgs.

\clearpage

{\bf UPCs as the energy frontier} Since the closure of the HERA $ep$ collider, there have been no dedicated high-energy photon facilities.  Instead, the photon energy frontier has been studied in UPCs at CERN's Large Hadron Collider \cite{Baur:2001jj,Greiner:1992fz,Bertulani:2005ru,Baltz:2007kq,Contreras:2015dqa,Klein:2017nqo,Klein:2020fmr}.   The protons and heavy ions accelerated there carry electromagnetic fields which may be treated as a flux of nearly-real (virtuality $Q^2< (\hbar c/R_A)^2$, where $R_A$ is the hadron radius) photons.  The photon spectra extend up to energies of $\gamma\hbar c/R_A$, where $\gamma$ is the Lorentz boost of the ion. At the LHC, these photons lead to $\gamma p$ collisions at center of mass energies up to 5 TeV, $\gamma A$ collisions at center of mass energies up to 700 GeV/nucleon, and two-photon collisions up to $\sqrt{s_{\gamma\gamma}} = 4.2$ TeV.  The $\gamma p$ energies are higher than were accessible at HERA, and the $\gamma A$ energies are many orders of magnitude higher than are accessible at fixed-target experiments. These photons have been used to study a wide variety of physics processes: measurements of low$-x$ gluon densities in protons and studies of shadowing of parton densities in heavier nuclei, studies of higher order terms in dilepton production, and light-by-light scattering.  

{\bf Photoproduction and parton distributions} Parton distributions have been probed in $\gamma p$ and $\gamma A$ collisions. 
Photoproduction of dijets (and, still to come, open charm \cite{Klein:2002wm}) is, from the theoretical point of view, a fairly direct probe of the gluon distribution.  So far, preliminary results from ATLAS on dijet production \cite{ATLAS:2017kwa} and from CMS on exclusive dijet photoproduction~\cite{CMS:2020ekd} have been released. This can be used to measure the diffractive structure functions and study ``elliptic gluon" dynamics~\cite{Guzey:2016tek,Kotko:2017oxg,Altinoluk:2015dpi,Hagiwara:2017fye,Mantysaari:2019csc,Hatta:2016dxp,Zhou:2016rnt}. UPCs at the LHC can probe to Bjorken$-x$ values of at least a few $10^{-6}$; this could reach even lower $x$ with far-forward detectors like the proposed ALICE FoCal \cite{vanderKolk:2020fqo}.

Exclusive vector mesons are produced when an incident photon fluctuates to a virtual quark-antiquark pair, which then scatters elastically (or quasi-elastically) from a proton or nuclear target.  Since elastic scattering requires two-gluon exchange for color neutrality, the cross section scales as the square of the gluon density.  One limitation is that the color neutrality requirement introduces some systematic uncertainty \cite{Flett:2019ept,Flett:2019pux}.    
ALICE \cite{Acharya:2018jua} and LHCb have studied $J/\psi$ production on proton targets, finding, for the most part, that the power-law behavior seen at HERA extends to higher energies. ALICE, LHCb and CMS have also studied $\psi(2S)$ \cite{Adam:2015sia,LHCb:2016oce}and $\Upsilon$ production \cite{Aaij:2015kea,McNulty:2016sor,Sirunyan:2018sav}, finding good agreement with NLO-inspired cross section calculations. Some calculations indicate that the saturation scale has been reached~\cite{Garcia:2019tne}.

The nuclear-target cross sections are sensitive to gluon shadowing, and  probe phenomena like gluon saturation of the color-glass condensate. ALICE \cite{Abbas:2013oua,Acharya:2019vlb} and CMS \cite{Khachatryan:2016qhq} have studied $J/\psi$ production on lead, and found moderate suppression, consistent with leading-twist calculations \cite{Guzey:2013xba,Klein:2019qfb,Guzey:2020ntc}.  There have also been studies on $\rho$ photoproduction \cite{Sirunyan:2019nog,Abelev:2007nb,Adam:2015gsa}.  
The $\rho $ cross section is about 40\% smaller than predicted by the Glauber model, pointing to the importance of high-mass internal states, as expected in the Glauber-Gribov model \cite{Frankfurt:2015cwa}. 

Looking ahead, the LHC and HL-LHC will collect much more data, which should lead to precise measurements of shadowing (the anticipated error bars are shown in Fig. 1) and enable new types of measurements. In the Good-Walker paradigm \cite{Good:1960ba}, coherent and incoherent photoproduction allow access to qualitatively new studies \cite{Mantysaari:2016jaz,Klein:2019qfb}. The coherent cross section $d\sigma/dt$ encodes the transverse spatial distribution of the targets - the nuclear equivalent of a Generalized Parton Distribution - while the incoherent $d\sigma/dt$ provides information on event-by-event fluctuations in the nuclear configuration, due to varying nucleon positions and gluonic hot spots \cite{Mantysaari:2016jaz,Guzey:2016qwo,Cepila:2016uku,Cepila:2018zky}. STAR has used $d\sigma/dt$ for
$\rho$ and direct $\pi^+\pi^-$ photoproduction to measure the spatial distribution of target scatters in gold \cite{Adamczyk:2017vfu}. HL-LHC can do this measurement with heavier quarkonia, where pQCD is clearly applicable. CMS data on $\rho$~\cite{Sirunyan:2019nog} shows sensitivity to the onset of gluon saturation in lead using differential studies in both momentum transfer ($t$) and $\gamma p$ center-of-mass energy~\cite{Goncalves:2018blz}. Other opportunities include study of perturbative Pomeron dynamics~\cite{Blok:2010ds}, color fluctuations in the photon~\cite{Alvioli:2016gfo,Frankfurt:2008et}, the gluonic Sivers function~\cite{Boussarie:2019vmk}, search for the Odderon\cite{Harland-Lang:2018ytk}, among other studies. Finally, next-to-leading (NLO) order calculations for these processes are one of the future directions of the theoretical program~\cite{Boussarie:2014lxa,Boussarie:2016ogo,Boussarie:2016bkq}. 

\begin{figure}[tb!]
  \vspace*{-0.5cm}
  \centering
  \includegraphics[width=0.4\textwidth]{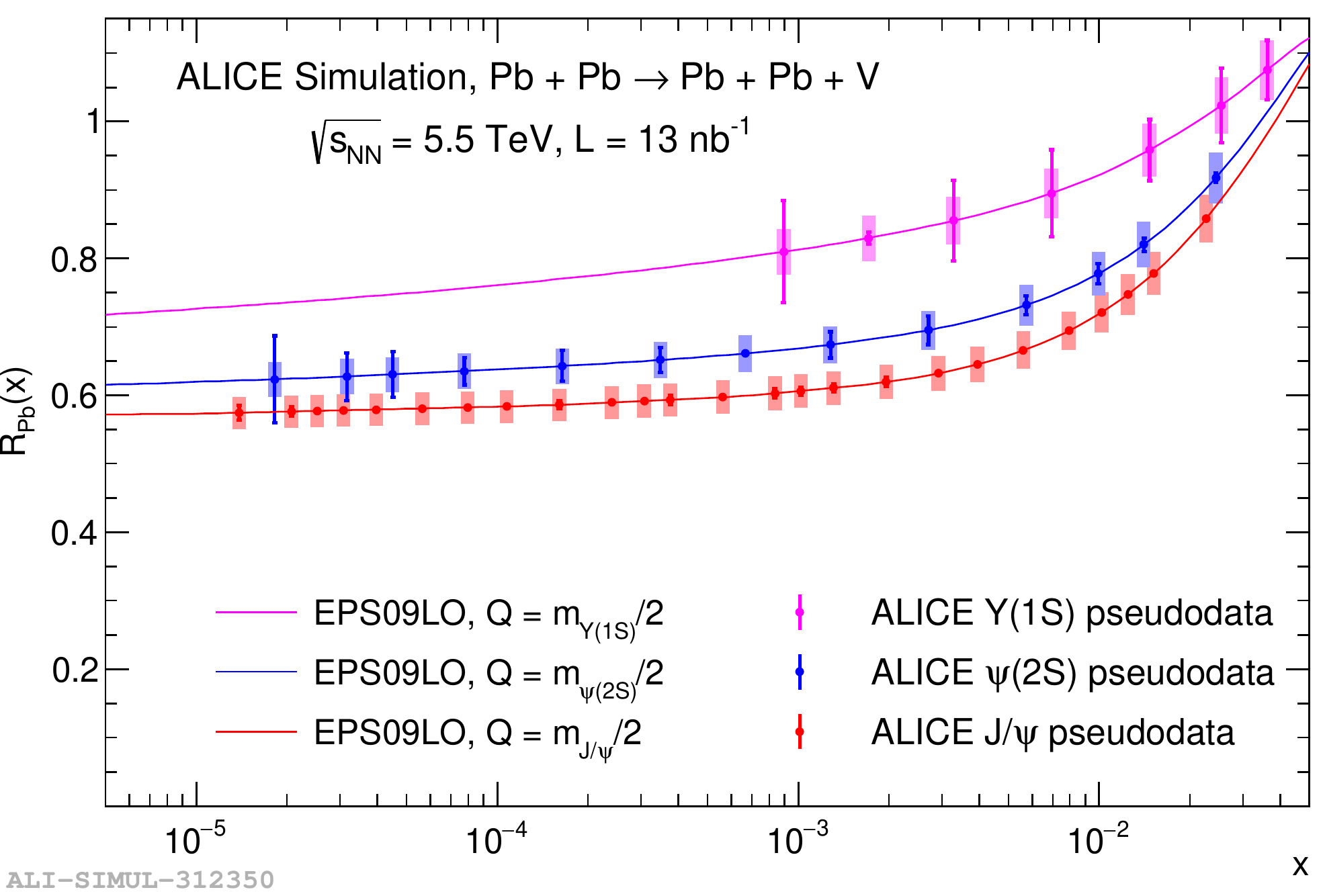}
  \includegraphics[width=0.44\textwidth]{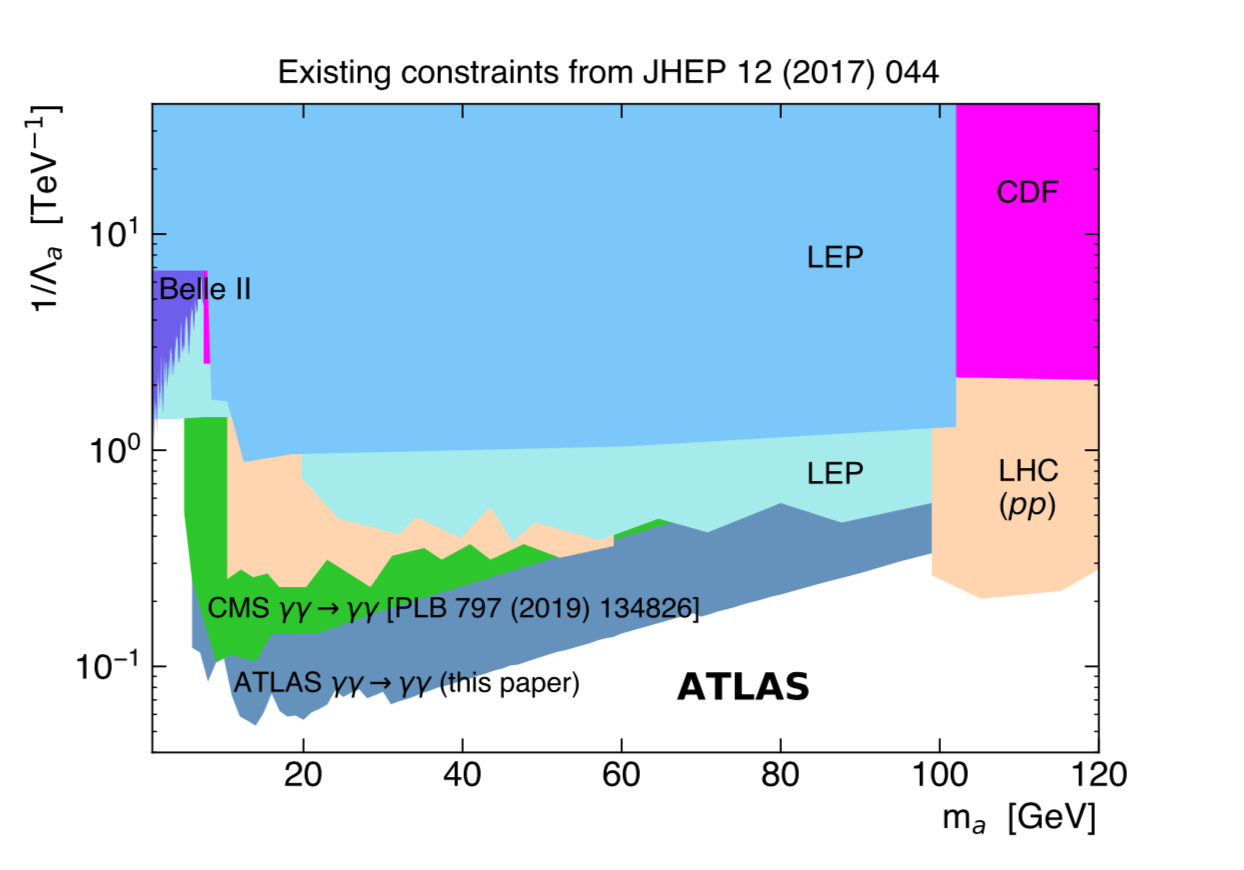}
  \caption{\label{fig:1} (left) Pseudodata projections for the nuclear suppression factors of the nuclear gluon density as a function of Bjorken-$x$ for photoproduction of three vector mesons in PbPb UPCs. The points are EPS09-projections using the method described in~\cite{Guzey:2013xba}. 
   From Ref. \cite{Citron:2018lsq}. (right) Exclusion limits on ALP-photon coupling ($1/\Lambda_a)$ vs. ALP mass, from light-by-light scattering and other processes.  From Ref. \cite{Aad:2020cje}.
 }
\end{figure}

{\bf Light-by-light scattering, $W$ pair and dilepton production.}  Two-photon interactions at the LHCb probe the energy frontier.  Photons couple to all electrically charged particles and some neutrals scalars (like axions and the Higgs), so two-photon reactions are sensitive to many beyond-standard-model processes.  

One process that has already been used to probe BSM physics is light-by-light scattering, $AA\rightarrow AA\gamma\gamma$ \cite{dEnterria:2013zqi,Coelho:2020syp}.  The subprocess $\gamma\gamma\rightarrow\gamma\gamma$ proceeds only via a charged-particle box diagram.  The cross section is sensitive to all charged particles, including BSM particles \cite{Fichet:2014uka,Bauer:2017ris} such as vector fermions, GeV- mass axion-like particles (ALPs) \cite{Knapen:2016moh,Coelho:2020saz} and magnetic monopoles.  The reaction also probes non-linear (BSM) corrections to electromagnetism \cite{Ellis:2017edi,Akmansoy:2018xvd}.  ATLAS \cite{Aaboud:2017bwk,Aad:2019ock,Aad:2020cje} and CMS \cite{Sirunyan:2018fhl} have both observed this process at a cross section consistent with the standard model.  They then set limits on ALP production, as shown in Fig. \ref{fig:1}.

In pp collisions, ATLAS \cite{ATLAS:2020qfn} and CMS \cite{Chatrchyan:2013foa} have also studied $\gamma\gamma\rightarrow W^+W^-$ and CMS/TOTEM $\gamma\gamma\rightarrow \gamma\gamma$\cite{CMS:2020rzi}, thereby placing stringent limits on anomalous quartic gauge couplings.  Production of electron and muon pairs via $\gamma\gamma\rightarrow l^+l^-$, has been studied by many collaborations. Exclusive $e-\mu$ events, as expected from tau pairs \cite{ATLAS-EVENTDISPLAY-2018-009} have been seen.  Tau pairs are of particular interest because the $\tau$-$\gamma$ coupling is sensitive to BSM physics, including the $\tau$ dipole moment, lepton compositeness or supersymmetry.  

Looking ahead, $\gamma\gamma\rightarrow\gamma\gamma$ is a rare process and future HL-LHC running should lead to much larger data samples for this final states \cite{Citron:2018lsq}. ALICE and LHCb may be able to probe this reaction at lower diphoton masses than is currently possible~\cite{Klusek-Gawenda:2019ijn}.  High statistics studies of dilepton production may also be sensitive to BSM processes, particularly at high masses.  This is especially true for the $\tau$, where the cross section \cite{Beresford:2019gww} and 
kinematic distributions \cite{Dyndal:2020yen} are sensitive to new physics. It may also be possible to study two-photon production of heavy quark states, $\gamma \gamma\rightarrow p\bar{p}$~\cite{Klusek-Gawenda:2017lgt}, and search for pentaquarks~\cite{Goncalves:2019vvo}, tetraquarks \cite{Klein:2019avl} and other exotica.

{\bf Strong fields, Quantum correlations and quantum tomography}
UPCs also offer opportunities to use the very strong fields to explore reactions involving multiple photon exchange, such as production of $\rho^0\rho^0$, where each $\rho^0$ is produced by a single, independent photon \cite{Klein:1999qj,Klusek-Gawenda:2013dka,Goncalves:2016ybl}. These pairs should exhibit quantum correlations which will allow for a more detailed study of decay dynamics, including with polarized photons. EPR (Einstein-Podolsky-Rosen)-type experiments can also be carried out to test quantum mechanics, and quantum tomography techniques can probe quantum correlations and entanglement~\cite{Klein:2002gc,Martens:2017cvj}.  

{\bf UPCs at the FCC and synergies with future colliders}
Photon exchange at the future FCC, or at the proposed LHeC~\cite{Agostini:2020fmq}, will allow probes at even higher energies, allowing for more extensive BSM physics and also the study of top photoproduction \cite{Klein:2000dk,Goncalves:2013oga} and $\gamma\gamma$ production of the Higgs \cite{dEnterria:2019jty}.  The future electron-ion collider \cite{Accardi:2012qut}, planned to be built at BNL around 2030, will make precise measurements of photon-mediated reactions over a wide $Q^2$ range, but at lower energies.

\clearpage


\bibliographystyle{utphys}
\bibliography{photons}
\vspace{3.5in}


\end{document}